\documentclass[12pt,letterpaper]{article}
\usepackage{jheppub}

\title{The Cosmological Constant Problem, Dark Energy, and the Landscape of String Theory}

\author[a,b]{Raphael Bousso}
\affiliation[a]{Center for Theoretical Physics and Department of Physics,\\
 University of California, Berkeley, CA 94720, U.S.A.}
\affiliation[b]{Lawrence Berkeley National Laboratory, Berkeley, CA 94720,
  U.S.A.}

\abstract{% 
  In this colloquium-level account, I describe the cosmological constant problem: why is the energy of empty space at least 60 orders of magnitude smaller than several known contributions to it from the Standard Model of particle physics?  I explain why the ``dark energy'' responsible for the accelerated expansion of the universe is almost certainly vacuum energy.  The second half of the paper explores a more speculative subject.  The vacuum landscape of string theory leads to a multiverse in which many different three-dimensional vacua coexist, albeit in widely separated regions.  This can explain both the smallness of the observed vacuum energy and the coincidence that its magnitude is comparable to the present matter density.
}

\begin{document}
\maketitle

\section{The Cosmological Constant Problem}
\label{sec-ccp}

\subsection{A Classical Ambiguity}
\label{sec-classical}

In the field equation for General Relativity, 
\begin{equation}
  R_{\mu\nu} - \frac{1}{2} R g_{\mu\nu} + \Lambda g_{\mu\nu} =
  8\pi G T_{\mu\nu}~,
\label{eq-gr}
\end{equation}
there is an ambiguity: the cosmological constant, $\Lambda$, is not fixed by
the structure of the theory.\footnote{This paper aims at a level that would be accessible to a graduate student.  It is based on colloquia given at Caltech, MIT, and the University of Michigan, Ann Arbor, and on a lecture presented at {\em Subnuclear Physics: Past, Present and Future}, Pontificial Academy of Sciences, Vatican (October 2011, to appear in the proceedings).  In parts, I closely follow Refs.~\cite{Bou06b,TASI07}.}  There is no formal reason to set it to
zero, and in fact, Einstein famously tuned it to yield an (unstable) static
cosmological solution---his ``greatest blunder''.

After Hubble's discovery that the universe is expanding, the cosmological term was widely abandoned.  But setting $\Lambda=0$ was never particularly satisfying, even from a classical perspective.  The situation is similar to a famous shortcoming of Newtonian gravity: nothing prevents us from equating the gravitational charge with inertial mass, but nothing forces us to do so, either.

A nonzero value of $\Lambda$ introduces a length scale and time
scale
\begin{equation}
r_\Lambda= c t_\Lambda=\sqrt{3/|\Lambda|}
\label{eq-rlam}
\end{equation}
into General Relativity.  An independent length scale arises
from the constants of Nature: the Planck length\footnote{Here $G$
  denotes Newton's constant and $c$ is the speed of light. In this
  paper Planck units are used unless other units are given explicitly.
  For example, $t_{\rm P}=l_{\rm P} /c \approx .539 \times 10^{-43}
  {\rm s}$ and $M_{\rm P} = 2.177\times 10^{-5} {\rm g}$.}
\begin{equation}
  l_{\rm P} = \sqrt{\frac{G\hbar}{c^3}}
  \approx 1.616 \times 10^{-33} {\rm cm}~.
\end{equation}
It has long been known empirically that $\Lambda$ is very small in Planck units (i.e., that $r_\Lambda$ is large in these natural units).  This can be deduced just from the fact that the universe is large compared to the Planck length, and old compared to the Planck time.

First, consider the case of positive $\Lambda$.  If no matter is present ($T_{\mu\nu}=0$), then the only isotropic solution to Einstein's equation is de~Sitter space, which exhibits a cosmological horizon of radius $r_\Lambda$~\cite{HawEll}.  A cosmological horizon is the largest observable distance scale, and the presence of matter will only decrease the horizon radius~\cite{GibHaw77a,Bou00a}.  But we observe scales that are large in Planck units ($r\gg 1$).  Since $r_\Lambda$ must be even larger, Eq.~(\ref{eq-rlam}) implies that the cosmological constant is small.

Negative $\Lambda$ causes the universe to recollapse independently of spatial curvature, on a timescale $t_{\Lambda}$~\cite{Edw72}.  Thus, the large age of the universe (in Planck units) implies that $(-\Lambda)$ is small.  Summarizing the above arguments, one finds \begin{equation} -3t^{-2}\lesssim\Lambda\lesssim 3r^{-2}~, \label{eq-trlam} \end{equation} 
where $t$ and $r$ are any time scale and any distance scale that have been observed. We can see out to distances of billions of light years, so $r>10^{60}$; and stars are billions of years old, so $t>10^{60}$.  With these data, known for many decades, Eq.~(\ref{eq-trlam}) implies roughly that \begin{equation}
  |\Lambda|\lesssim 3\times 10^{-120}~.
\label{eq-small}
\end{equation}
Thus, in Planck units, $\Lambda$ is very small indeed.

This result makes it tempting to set $\Lambda=0$ in the Einstein
equation; and at the level of the classical gravity theory, we are free to do so.  However, in Eq.~(\ref{eq-gr}), the $\Lambda$-term is not the only term proportional to the metric.  Another, much more problematic contribution enters through the stress tensor on the right hand side.

\subsection{Quantum Contributions to $\Lambda$}
\label{sec-quantum}

In quantum field theory, the vacuum is highly nontrivial.\footnote{Further details can be found in Weinberg's classic review~\cite{Wei89}.  Among more recent reviews, I recommend Polchinski's concise discussion of the cosmological constant problem~\cite{Pol06}, which I follow in parts of this subsection.}  In the Standard Model, the vacuum is responsible for physical phenomena such as confinement and the Higgs mechanism.  Like any physical object, the vacuum will have an energy density.  
Lorentz invariance requires that the corresponding energy-momentum-stress tensor be proportional to the metric,
\begin{equation}
\langle T_{\mu\nu} \rangle = -\rho_{\rm vacuum} g_{\mu\nu} ~.
\label{eq-lrho}
\end{equation}
This is confirmed by direct calculation. (See any introductory textbook on quantum field theory, such as Ref.~\cite{Zee}.)  The form of the stress tensor ensures that the vacuum looks the same to all observers independently of orientation or velocity.  This property (and not, for example, vanishing energy density) is what distinguishes the vacuum from other objects such as a table.

Though it appears on the right hand side of Einstein's equation,
vacuum energy has the form of a cosmological constant, and one might as
well absorb it and redefine $\Lambda$ via
\begin{equation}
  \Lambda=\Lambda_{\rm Einstein}   +8\pi\rho_{\rm vacuum}~.
\end{equation}
Equivalently, one may absorb the ``bare'' cosmological constant
appearing in Einstein's equation, $\Lambda_{\rm Einstein}$, into the energy
density of the vacuum, defining
\begin{equation}
\rho_\Lambda\equiv \rho_{\rm vacuum}+\frac{\Lambda_{\rm Einstein}}{8\pi}~.
\end{equation}

\begin{figure}[tbp]
\centering
\includegraphics[width=.8\textwidth]{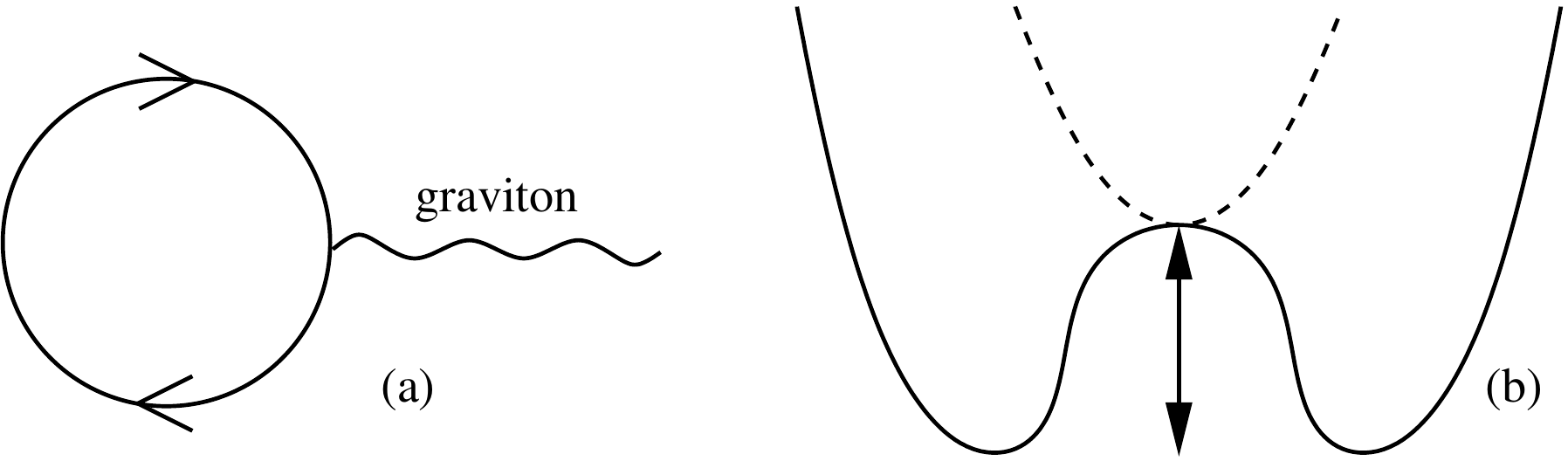}
 \caption{Perturbative and nonperturbative
  contributions to vacuum energy.  (a) Virtual particle-antiparticle
  pairs, the zero-point fluctuations of the quantum fields (b) Effective scalar field potentials, such as the potential for the Higgs field shown here schematically.  Before electroweak symmetry
  breaking in the early universe the vacuum energy was about 56 orders of magnitude greater than todays value (dashed line).}
\label{fig-loop} 
\end{figure}
Eqs.~(\ref{eq-rlam}), (\ref{eq-trlam}), and (\ref{eq-small}) apply to
the total cosmological constant, and can be restated as an empirical
bound on the total energy density of the vacuum:
\begin{equation}
  |\rho_\Lambda|\lesssim 10^{-121}~.
\end{equation}
But in the Standard Model, the energy of the vacuum receives many
contributions much larger than this bound.  Their value depends on the
energy scale up to which we trust the theory.  It is enormous even with
a conservative cutoff.  

This would be true already in free field theory.  Like a harmonic oscillator in the ground state, every mode of every free field contributes a zero-point energy to the energy density of the vacuum.  In a path integral description, this energy arises from virtual particle-antiparticle pairs, or ``loops'' (Fig.~\ref{fig-loop}a).  For example, consider the electron, which is well understood at least up to energies of order $M=$ 100 GeV~\cite{Pol06}.  Dimensional analysis implies that electron loops up to this cutoff contribute of order $(100$ GeV$)^4$ to the vacuum energy, or $10^{-68}$ in Planck units.  

Similar contributions are expected from other fields and from interactions.  The real cutoff is probably of order the supersymmetry breaking scale, giving at least $(1$ TeV$)^4\approx 10^{-64}$.  It may be as high as the Planck scale, which would yield $|\rho_\Lambda|$ of order unity.\footnote{Recall that Planck units are used throughout.  $\rho_\Lambda=1$ would correspond to a density of $10^{94}$ g/cm$^3$.}  Thus, quantum field theory predicts multiple perturbative contributions to $|\rho_\Lambda|$.  Each contribution is some 60 to 120 orders of magnitude larger than the experimental bound, Eq.~(\ref{eq-small}).

Additional contributions come from the effective potentials of scalar fields, such as the potential giving rise to symmetry breaking in the electroweak theory (Fig.~\ref{fig-loop}b).  The vacuum energy of the symmetric and the broken phase differ by approximately $(200$ GeV$)^4\approx 10^{-67}$.  Other symmetry breaking mechanisms at higher or lower energy, such as chiral symmetry breaking of QCD with $(300$ MeV$)^4\approx 10^{-79}$, will also contribute.  There is no reason why the total vacuum energy should be small in the symmetric phase, and even less so in the broken phase that the universe is in now.

I have exhibited various known contributions to the vacuum energy.
They are uncorrelated with one another and with the (unknown) bare
cosmological constant appearing in Einstein's equation, $\Lambda_{\rm
  Einstein}$.  Each contribution is dozens of orders of magnitude
larger than the empirical bound today, Eq.~(\ref{eq-small}).  In
particular, the radiative correction terms from quantum fields are
expected to be at least of order $10^{-64}$.  They come with different
signs, but it would seem overwhelmingly unlikely for such large terms
to cancel to better than a part in $10^{120}$, in the present era.

This is the cosmological constant problem: {\em Why is the vacuum
  energy today so small?\/}   It represents a serious crisis in physics: a
discrepancy between theory and experiment, of 60 to 120 orders of
magnitude.  What makes this problem hard is that it arises from two otherwise extremely successful theories---the Standard Model and General Relativity---in a regime where both theories have been reliably and precisely tested and hence cannot be dramatically modified.

\section{The Cosmological Constant}
\label{sec-duck}

In exhibiting the cosmological constant problem, I made use only of a
rather crude, and old, upper bound on the magnitude of the
cosmological constant.  The precise value of $\Lambda$ is irrelevant
as far as the cosmological constant {\em problem} is concerned: we
have known for several decades that $\Lambda$ is certainly much smaller than
typical contributions to the vacuum energy that can be estimated from
the Standard Model of particle physics.  In this section, I will discuss the observed value and its implications.

\subsection{Observed Value of $\Lambda$}

The actual value of $\Lambda$ was first determined in 1998 from the
apparent luminosity of distant supernovae~\cite{Rie98,Per98}.  Their
dimness indicates that the expansion of the universe has recently
begun to accelerate, consistent with a positive cosmological
constant
\begin{equation}
  \rho_\Lambda= (1.35\pm 0.15) \times 10^{-123}~,
\label{eq-duck}
\end{equation}
and inconsistent with $\rho_\Lambda =0$.  The quoted value and error bars are recent (WMAP7 + BAO + $H_0$~\cite{WMAP7}) and thus significantly improved relative to the original discovery.

Cross-checks have corroborated this conclusion.  For example, the above value of $\rho_\Lambda$ also explains the observed spatial flatness of the universe~\cite{WMAP7}, which cannot be accounted for by baryonic and dark matter alone.  And surveys of the history of structure formation in the universe~\cite{Rei09} reveal a recent disruption of hierarchical clustering consistent with accelerated expansion driven by the cosmological constant of Eq.~(\ref{eq-duck}).

\subsection{Why Dark Energy is Vacuum Energy}

The observed vacuum energy, Eq.~(\ref{eq-duck}), is sometimes referred to as ``dark energy''.  This choice of words is meant to be inclusive of other possible interpretations of the data, in which $\Lambda=0$.  Dark energy might be a form of scalar matter (quintessence) which mimics a fixed cosmological constant closely enough to be compatible with observation, but retains some time-dependence that could in principle be discovered if it lurks just beyond current limits.  Another frequently considered possibility is that General Relativity is modified at distances comparable to the size of the visible universe, so as to mimic a positive cosmological constant even though $\Lambda=0$.  In both cases, model parameters can be adjusted to lead to predictions for future experiments that differ from those of a fixed cosmological constant.

Consideration of these theoretical possibilities, however, is at best premature.  It conflicts with a basic tenet of science: adopt the simplest interpretation of the data, and complicate your model only if forced to by further observation.  

Scenarios like quintessence or modified gravity are uncalled for by data and solve no theoretical problem.\footnote{Some models have been claimed to address the coincidence problem described in  Sec.~\ref{sec-coincidence} below.  Aside from unsolved technical problems~\cite{Car00}, what would be the point of addressing the (relatively vague) coincidence problem with a model that ignores the logically prior and far more severe cosmological constant problem (Sec.~\ref{sec-quantum})?}  In particular, they do not address the cosmological constant problem.  But such models contain adjustable parameters in addition to $\Lambda$.  Therefore, they are less predictive than the standard $\Lambda$CDM model.  Worse, in phenomenologically viable models, these additional parameters must be chosen small and fine-tuned in order to evade existing constraints.\footnote{For example, quintessence models require exceedingly flat scalar field potentials which must be fine-tuned against radiative corrections, and their interaction with other matter must be tuned small in order to be compatible with observational limits on a long-range fifth force~\cite{Car98,Car00}.  More natural models~\cite{HalNom05} have become difficult to reconcile with observational constraints.}  Again, such tunings are strictly {\em in addition} to the tuning of the the cosmological constant, which must be set to an unnaturally small or zero value in any case.

Therefore, dynamical dark energy should not be considered on the same footing with a pure cosmological constant.  The discovery of any deviation from a cosmological constant in future experiments is highly unlikely, as is the discovery of a modification to General Relativity on large scales.

A frequent misconception that appears to underlie the consideration of ``alternatives'' to $\Lambda$ is the notion that vacuum energy is somehow optional.  The idea is that the cosmological constant problem only arises if we ``assume'' that vacuum energy exists in the first place.  (This flawed argument is found in surprisingly prominent places~\cite{DETF}.)  It would be wonderful indeed if we could solve the cosmological constant problem with a single stroke, by declaring that vacuum energy just does not exist and setting $\Lambda$ to zero.

But in fact, we know that vacuum energy exists in Nature.  We can manipulate the amount of vacuum energy in bounded regions, in Casimir-type experiments.  And if $\Lambda$ had turned out to be unobservably small today, we would still know that it was large and positive in the early universe before electroweak symmetry breaking, according to the Standard Model of particle physics.\footnote{The theory of electroweak symmetry breaking is supported by overwhelming experimental evidence (chiefly, the $W$ and $Z$ bosons, and soon perhaps the Higgs).  It allows us to compute that $\Lambda\sim (200$ GeV$)^4$ at sufficiently high temperatures, when electroweak symmetry is unbroken~\cite{Pol06}.  Aside from the early universe, small regions with unbroken symmetry could be created in the laboratory, at least in principle.}  More generally, the notion that the vacuum has energy is inseparable from the experimental success of the Standard Model as a local quantum field theory~\cite{Pol06}.  

Contributions to $\Lambda$ from Standard Model fields are large, so the most straightforward theoretical estimate of its magnitude fails.  But just because $\Lambda$ should be much larger than the observed value does not imply that it must be zero.  In fact, no known extension or modification of the Standard Model predicts that $\Lambda=0$ without violently conflicting with other observations (such as the facts that the universe is not empty, and that supersymmetry, if it exists, is broken).

Thus, the cosmological constant problem is present either way, whether we imagine that $\Lambda$ is small (which is consistent with data) or that $\Lambda=0$ (which is not, unless further considerable complications are introduced).  Dark energy is experimentally indistinguishable from vacuum energy, and definitely distinct from any other previously observed form of matter.  The only reasonable conclusion is that dark energy is vacuum energy, and that its density is given by Eq.~(\ref{eq-duck}).

\subsection{The Coincidence Problem}
\label{sec-coincidence}

The observed value of $\Lambda$ does raise an interesting question, usually referred to as the coincidence problem or ``why now'' problem.  Vacuum energy, or anything behaving like it (which includes all options still allowed by current data) does not redshift like matter.  In the past, vacuum energy was negligible, and in the far future, matter will be very dilute and vacuum energy will dominate completely.  The two can be comparable only in a particular epoch.  It is intriguing that this is the same epoch in which we are making the observation.  

Note that this apparent coincidence involves us, the observers, in its very definition.  This constrains possible explanations (other than those involving an actual coincidence). In the following section, I will outline a framework which can solve both the coincidence problem and the (far more severe) cosmological constant problem of Sec.~\ref{sec-quantum}.

\section{The Landscape of String Theory and the Multiverse}
\label{sec-theory}

The string landscape is the only theoretical framework I am aware of that can explain why $\Lambda$ is small without conflicting with other data.\footnote{For alternative classes of approaches to the cosmological constant problem, and the obstructions they face, see Refs.~\cite{Pol06,TASI07}.}  (It is worth stressing, however, that the ideas I am about to discuss are still speculative, unlike those of the previous two sections.)  The way in which string theory addresses the cosmological constant problem can be summarized as follows:
\begin{itemize}
\item Fundamentally, space is nine-dimensional.   There are many distinct ways (perhaps $10^{500}$) of turning nine-dimensional space into three-dimensional space by compactifying six dimensions.\footnote{Amazingly, this idea was anticipated by Sakharov~\cite{Sak84} before string theory became widely known.}
\item Distinct compactifications correspond to different three-dimensional metastable vacua with different amounts of vacuum energy. In a small fraction of vacua, the cosmological constant will be accidentally small.
\item All vacua are dynamically produced as large, widely separated regions in spacetime
\item Regions with $\Lambda\sim 1$ contain at most a few bits of information and thus no complex structures of any kind.  Therefore, observers find themselves in regions with $\Lambda\ll 1$.
\end{itemize}

\subsection{The Landscape of String Theory}
\label{sec-landscape}

String theory is naturally formulated in nine or ten spatial dimensions~\cite{GSW,Polchinski}.  This does not contradict observation but implies that all but three of these dimensions are (effectively) compact and small, so that they would not have been observed in high-energy experiments.   I will discuss the case of six compact extra dimensions for definiteness.

Simple examples of six-dimensional compact manifolds include the six-sphere and the six-dimensional torus.  A much larger class of manifolds are the Calabi-Yau spaces, which have a number of useful properties and have been extensively studied.  They are topologically complex, with hundreds of distinct cycles of various dimensions.  Cycles are higher-dimensional analogues of the handles of a torus.  A rubber band that wraps a handle cannot be removed, or wrapped around a different handle, without ripping it apart.  A more pertinent example are electrical field lines, which can wrap a one-cycle (such as one of the cycles on a two-dimensional torus). 

String theory contains a certain set of nonperturbative objects known as $D$-branes, which act as sources of $D+2$ flux.  For example, a zero-brane is a pointlike object and sources a Maxwell field, much like an electron would.  Higher-dimensional objects such as membranes act as sources of higher-dimensional analogues of the Maxwell field.  Unlike in the Standard Model, however, the values of $D$ for which $D$-branes exist, their energy density, and their charge are all determined by consistency requirements.  They are set by the string scale and are not adjustible parameters.

$D$-branes and their associated fluxes can wrap topological cycles the same way that rubber bands and electric field lines can wrap the handles of a torus.  In string theory, the shape and size of the compact extra dimensions is determined by (among other things) the fluxes that wrap around the various topological cycles.  The geometry of spacetime is dynamical and governed by equations that limit to Einstein's equations in the appropriate limit.  The presence of matter will deform the compact manifold correspondingly; in particular, one expects that each cycle can at most support a few units of flux before gravitational backreaction causes it to pinch off (changing the topology of the compact manifold) or grow to infinite size (``decompactify'').  

Based on these arguments, we may suppose that there are on the order of 500 cycles, and that each can support between 0 and 9 units of flux.  Then there are $10^{500}$ different, distinct choices for the matter content, shape, and size of the extra dimensions.  This argument is a vast oversimplification, but it helps clarify how numbers like $10^{500}$ arise: by exponentiation of the number of topological cycles in a typical six-dimensional compact manifold.\footnote{For a more detailed nontechnical version of this argument, see Ref.~\cite{BouPol04}. Despite early results that the number of compactifications could be large~\cite{LerLus87}, the significance of this possibility was obscured by the unsolved problem of moduli stabilization and supersymmetry breaking~\cite{DouKac06}; see, however, Ref.~\cite{Sch98}.  The argument that string theory contains sufficiently many metastable vacua to solve the cosmological constant problem, and that vacua with $\Lambda\sim 10^{-123}$ are cosmologically produced and reheated was presented in Ref.~\cite{BP}.  An explicit construction of a large class of nonsupersymmetric flux vacua was first proposed in Ref.~\cite{KKLT}. (Constructions in noncritical string theory were
proposed earlier~\cite{Sil01,MalSil02}.)  More advanced counting
methods~\cite{DenDou04b} bear out the quantitative estimates of Ref.~\cite{BP} for the number of flux vacua.  See Ref.~\cite{DouKac06} for a review of flux vacua and further references.}

A useful way of picturing the set of three-dimensional vacua of string theory is as a potential function in a 500-dimensional discrete parameter space.  (Of course, as far as actual pictures go, two parameters will have to suffice, as in a real landscape.) Each metastable configuration of fluxes corresponds to a local minimum in the landscape.   In any one-dimensional cross-section of the parameter space, there will only be a handful of minima, but overall the number of minima can be of order $10^{500}$.

\subsection{The Spectrum of $\Lambda$}
\label{sec-spectrum}

\begin{figure}[tbp]
\centering
\includegraphics[width=.2\textwidth]{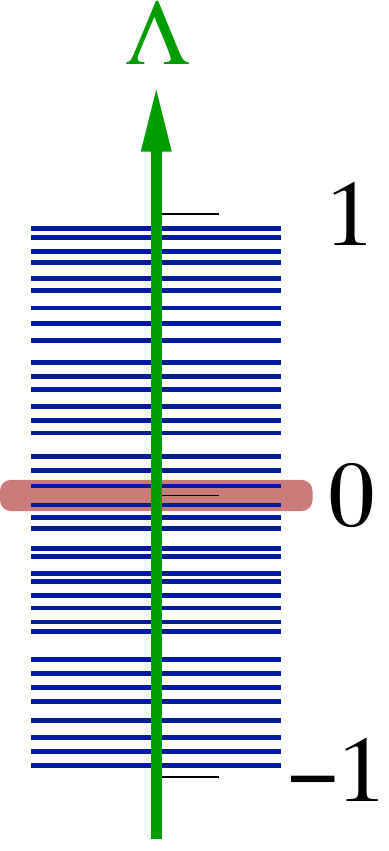}
 \caption{The spectrum of the cosmological constant (vacuum energy, dark energy) in the string landscape (schematic).  Each blue line represents one three-dimensional vacuum. With $10^{500}$ vacua, the spectrum will be very dense, and many vacua will have values of $\Lambda$ compatible with observation (red/shaded region).}
\label{fig-spec3} 
\end{figure}

Each vacuum has distinct matter and field content at low energies, determined by the matter content of the extra dimensions.  (Pictorially, the field spectrum corresponds to the details of each valley's shape near the minimum.)  In particular, the energy of each vacuum is essentially a random variable that receives positive and negative contributions from all particle species.  If we select one vacuum completely at random, the arguments of Sec.~\ref{sec-quantum} tell us that its cosmological constant will probably be large, presumably of order unity in Planck units (Fig.~\ref{fig-spec3})---as if we had thrown a dart at the interval $(-1,1)$, with an accuracy not much better than $\pm 1$.  

But this is true for every vacuum, so the overall spectrum of $\Lambda$ will be quite dense, with an average spacing of order $10^{-500}$.  This means that there will be a small fraction ($10^{-123}$) but a large number ($10^{377}$, in this example) of vacua with cosmological constant $|\Lambda|\lesssim 10^{-123}$.  Given enough darts, even a poor player will eventually hit the bullseye.

This is progress: at least, the theory contains vacua whose cosmological constant is compatible with observation.  But why is the universe in such a special, rare vacuum?  Did the universe start out in this particular valley of the landscape at the big bang, and if so, why?  In fact, there is no need to assume that initial conditions selected for a vacuum with small cosmological constant.  As we shall now see, such vacua are dynamically produced during cosmological evolution.

\subsection{de~Sitter Expansion and Vacuum Decay}

Suppose that the universe began in some vacuum with $\Lambda>0$.  Since about half of all vacua have positive energy, this is not a strong restriction.  We will not assume that the initial vacuum energy is particularly small; it may be of order one in Planck units.

The universe evolves as de~Sitter space, with metric
\begin{equation}
ds^2=-dt^2+e^{2Ht}(dr^2+r^2 d\Omega_2^2)~,
\label{eq-desitter}
\end{equation}
where the Hubble constant $H$ is given by $(\Lambda/3)^{1/2}$, and $d\Omega_2^2$ denotes the metric on the unit two-sphere.  This is an exponentially expanding homogeneous and isotropic cosmology.  In the following, it is not important that the universe looks globally like Eq.~(\ref{eq-desitter}).  It suffices to have a finite initial region larger than one horizon volume, of proper radius $e^{Ht_0} r> H^{-1}$.

Classically, this evolution would continue eternally, and no other vacua would ever come into existence anywhere in the universe.  This is because the vacuum itself is set by topological configurations of fluxes in the extra dimensions, which cannot change by classical evolution.  Quantum mechanically, however, it is possible for fluxes to change by discrete amounts.  This happens by a process completely analogous to the Schwinger process.  

The Schwinger process is the spontaneous pair production of electrons and positrons in a strong electric field between two capacitor plates.  It can be treated as a tunneling process in the semi-classical approximation.  The two particles appear at a distance at which the part of the field that their charges cancel out compensates for their total rest mass, so that energy is conserved.  Then the particles move apart with constant acceleration, driven by the remaining electric field, until they hit the plates (or in the case where the field lines wrap a topological circle, until they hit each other).  The final result is that the electric flux has been lowered by a discrete amount, corresponding to removing one unit of electric charge from each capacitor plate.

Similarly, the amount of flux in the six extra dimensions can change as a result of Schwinger-like processes, whereby branes of appropriate dimension are spontaneously nucleated.  (The Schwinger process itself is recovered in the case of zero-branes, i.e., charged point particles.)  Again, this is a nonperturbative tunneling effect.  Its rate is suppressed by the exponential of the brane action and is generically exponentially small.

Let us now give a description of this process from the 3+1 dimensional viewpoint. The effect of the six extra dimensions is to provide an effective potential landscape.  Each minimum corresponds to a metastable vacuum with three large spatial dimensions.   (Recall that the hundreds of dimensions of the landscape itself correspond to the topological cycles of the extra dimensions, not to actual spatial directions.)

The decay of a unit of flux, in this picture, corresponds to a transition from a higher to a lower-energy minimum in the potential landscape of string theory.\footnote{The following description of vacuum decay is a straightforward application of seminal results of Coleman for a one-dimensional potential with two vacua~\cite{Col77,CDL}. More complicated decay channels can arise in multidimensional potentials~\cite{BroDah10}; they do not affect the conclusions presented here.}  This transition does not happen simultaneously everywhere in three-dimensional space, because that process would have infinite action.  Rather, a bubble of the new vacuum appears spontaneously, as in a first-order phase transition.  Like in the Schwinger process, the initial size of the bubble is controlled by energy conservation.  The bubble wall is a domain wall that interpolates between two vacua in the effective potential.  The gradient and potential energy in the domain wall are compensated by the vacuum energy difference in the enclosed volume.

The bubble expands at constant acceleration.  As it moves outward, it converts the old, higher energy parent vacuum into a new, lower-energy vacuum.  The vacuum energy difference pays not only for the ever-expanding domain wall but can also lead to the production of matter and radiation inside the new vacuum.

The symmetries of a first-order phase transition in a relativistic theory dictate that the region inside the bubble is an open (i.e., negatively curved) Friedmann-Robertson-Walker universe.  In particular, time slices of constant density are infinitely large, even though the bubble starts out at finite size.  (This is possible because the choice of time variable in which we see the bubble expand is different from, and indeed inconsistent with, a choice in which constant time corresponds to hypersurfaces of constant density within the bubble.)  For this reason, the interior of the bubble is sometimes referred to as a ``universe'', ``pocket universe'', or ``bubble universe'', even though it does not constitute all of the global spacetime.

\subsection{Eternal Inflation}

We now turn to a crucial aspect of the decay of a metastable vacuum with positive energy: despite the decay and the expansion of the daughter bubble, the parent vacuum persists indefinitely.  This effect is known as eternal inflation~\cite{GutWei83,Lin86a}.  

The volume occupied by the parent vacuum expands exponentially at a rate set by its own Hubble scale $3H=3(3/\Lambda)^{1/2}$.  Some volume is lost to decay, at a rate $\Gamma$ per unit Hubble volume.  As long as $\Gamma\ll 3H$ (which is generic due to the exponentially suppressed nature of vacuum decay), the exponential expansion wins out, and the parent vacuum region grows on average.

The fact that the new vacuum expands after it first appears does not affect this result, since different regions in de~Sitter space are shielded from one another by cosmological event horizons.  A straightforward analysis of light propagation in the metric of Eq.~(\ref{eq-desitter}) shows that any observer (represented by a timelike geodesic) is surrounded by a horizon of radius $H^-1$.  The observer cannot receive any signals from any point $p$ beyond this horizon, by causality, no matter how long they wait.  A bubble of a new vacuum that forms at $p$ cannot expand faster than the speed of light (though it does expand practically at that speed).  Therefore it can never reach an observer who is initially more than a distance $H^{-1}$ from $p$ at the time of bubble nucleation.

Because the parent vacuum continues to grow in volume, it will decay not once but infinitely many times.  Infinitely many bubble universes will be spawned; yet, the overall volume of parent vacuum will continue to increase at a rate set by $3H-\Gamma\approx 3H$.    If the parent vacuum has multiple decay channels, then each decay type will be realized infinitely many times.  For example, in the string landscape we expect that a de Sitter vacuum can decay to any one of its hundreds of immediate neighbor vacua in the high-dimensional potential landscape.   All of these vacua will actually be produced as bubble universes, in exponentially distant regions, over and over.

\subsection{The Multiverse}

Let us now turn our attention to one of the daughter universes.  It is useful to distinguish three cases, according to the sign of its cosmological constant.  First, suppose that its vacuum energy is positive and that the vacuum is sufficiently long-lived (greater than about $t_\Lambda$).  In this case, the daughter universe will enter a phase of exponential de~Sitter expansion, beginning at a time of order $t_\Lambda$ after its nucleation.  It will give rise to eternal inflation in its own right, decaying in infinitely many places and producing daughter universes, while persisting globally.

\begin{figure}[tbp]
\centering
\includegraphics[width=.8\textwidth]{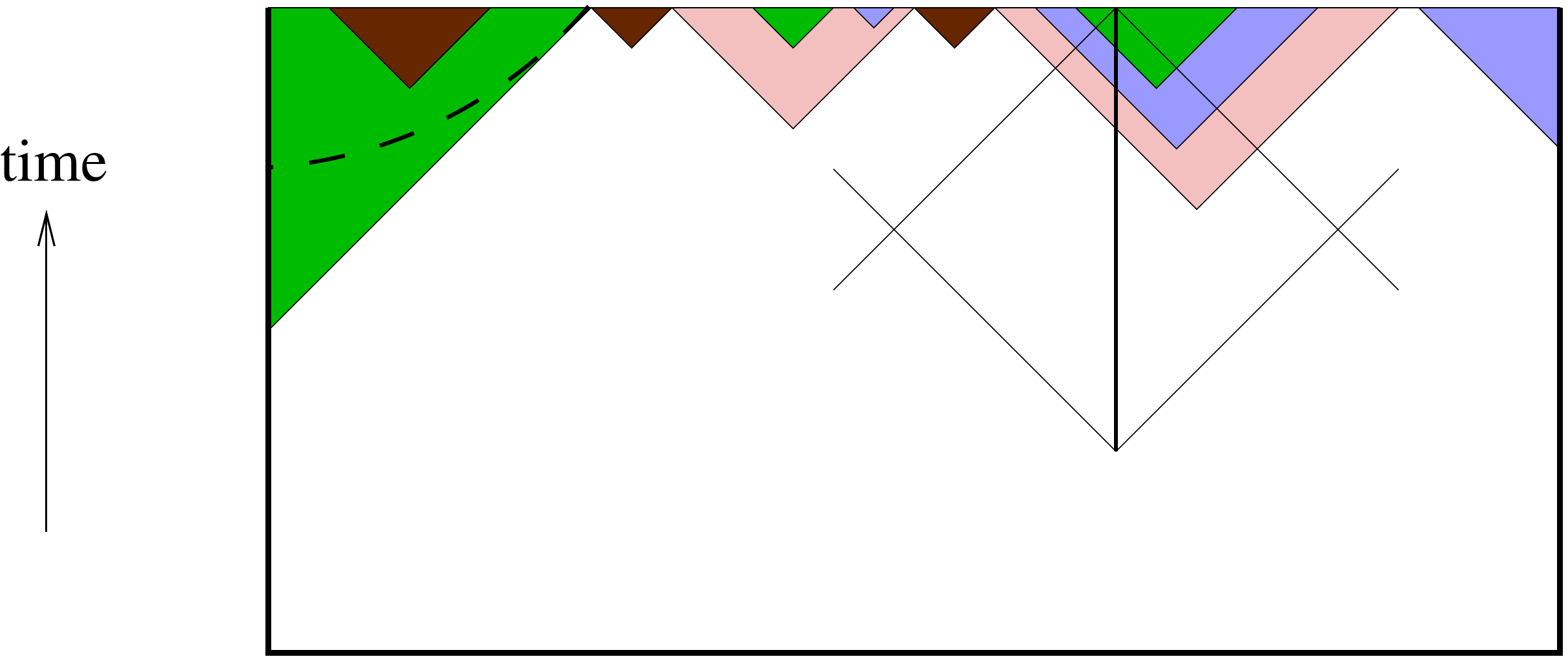}
 \caption{Conformal diagram of an eternally expanding multiverse (schematic).  Light travels at 45 degrees.  Different colors/shades represent different vacua in the string landscape.  Bubble universes have a triangular shape in this diagram.  They are bounded by domain walls whose expansion is so rapid that they look like future light-cones.  Event horizons shield different regions from one another: a hypothetical observer who survives multiple vacuum decays (black vertical line) would still only be able to probe a finite region in the infinite multiverse (black diamond).}
\label{fig-global} 
\end{figure}
Thus, the entire landscape of string theory can in principle be populated.  All vacua are produced dynamically, in widely separated regions of spacetime, and each is produced infinitely many times. This can be illustrated in a conformal diagram (or ``Penrose diagram''), which rescales the spacetime metric to render it finite but preserves causal relations (Fig.~\ref{fig-global}).  By convention, light-rays propagate at 45 degrees.  Bubbles look like future light-cones because they expand nearly at the speed of light.  Bubble universes that form at late times are shown small due to the rescaling, even though their physical properties are independent of the time of their production.  As a result of eternal inflation, the future boundary of the diagram has a fractal structure. 

Vacua with nonpositive cosmological constant are ``terminal''.  They do not give rise to eternal inflation.  If $\Lambda<0$, then the bubble universe begins to contract and collapses in a big crunch on a timescale of order $t_\Lambda$~\cite{CDL}.  The spacelike singularity does not reach outside the bubble universe with $\Lambda<0$; it does not affect global eternal inflation.  

One expects that the case $\Lambda=0$ arises only in vacua with unbroken supersymmetry.  They are completely stable and do not end in a crunch.  In the conformal diagram, they correspond to the ``hat regions'' near the future boundary (not shown in Fig.~\ref{fig-global}).

\subsection{Why Observers are Located in Regions With $|\Lambda|\ll 1$}

I have argued that the string landscape contains vacua with very small cosmological constant, such as ours.  Moreover, such vacua will be dynamically produced by inflation, starting from generic initial conditions.  But the bubble universes with $|\Lambda\ll 1$, such as ours, are surely very atypical regions in the large multiverse.  Typical regions (by almost any conceivable definition of ``typical'') would have cosmological constant of order one in Planck units, since almost all vacua have this property.  Why, then, do we find ourselves in one of the rare locations with $\Lambda \ll 1$?  

Before addressing this question, it is worth noting that the same question could not be asked in a theory that failed to contain vacua with $\Lambda\ll 1$, or that failed to produce such vacua as spacetime regions.  But in a theory that dynamically produces highly variable environments in different locations, it is important to understand correlations between environmental properties and the location of observers.  What is typically observed depends on where one is observing, so these correlations will affect the predictions of the theory.

In Sec.~\ref{sec-ccp}, I discussed that the cosmological constant sets a largest observable length or time scale, of order $|\Lambda^{-1/2}|$.  A more precise result can be stated in terms of the maximum area on the past light-cone of an arbitrary point (event) $p$ in a universe with nonzero cosmological constant~\cite{BouFre10a}.  If $\Lambda>0$, the past light-cone of any point $p$ has maximum area of order $\Lambda^{-1}$; if $\Lambda<0$, it has maximum area of order $|\Lambda|^{-1}$ (if the universe is spatially flat), or $\Lambda^{-2}$ (if the universe is open).  

The maximum area on the past light-cone of $p$, in units of the Planck length squared, is an upper bound on the entropy in the causal past of $p$:
\begin{equation}
S\lesssim A
\end{equation}
This follows from the covariant entropy bound~\cite{CEB1,RMP}.  It implies that regions with $\Lambda\sim 1$ do not contain more than a few bits of information in any causally connected region.  Whatever observers are made of, they presumably require more than one or two particles.  

This means that observers can only be located in regions with $|\Lambda|\ll 1$.   Because of cosmological horizons, they will not typically be able to see other regions.  Though typical regions have $\Lambda=1$, observations are made in regions with $|\Lambda|\ll 1$.  

\subsection{Predicted Value of $\Lambda$}

The argument shows only that $|\Lambda|\ll 1$ is a prediction of the string landscape; it does not explain why we see the particular value $\Lambda\sim 10^{-123}$.  In order to make this, or any other quantitative prediction, we would need to begin by regulating the infinities of eternal inflation.  This is known as the ``measure problem'', and it has little to do with the string landscape.  

The measure problem arises in any theory that gives rise to eternal inflation.  For this, one long-lived metastable de~Sitter vacuum is enough.  We appear to live in such a vacuum, so the measure problem needs attention independently of the number of other vacua in the theory.  A discussion of this problem and of current approaches to its solution would go beyond the scope of the present paper.   The reader is referred to Ref.~\cite{BouFre10d} and references therein; here we quote only the main result of this paper (see also Ref.~\cite{BouHar07,BouHar10}).

\begin{figure}[tbp]
\centering
\includegraphics[width=.8\textwidth]{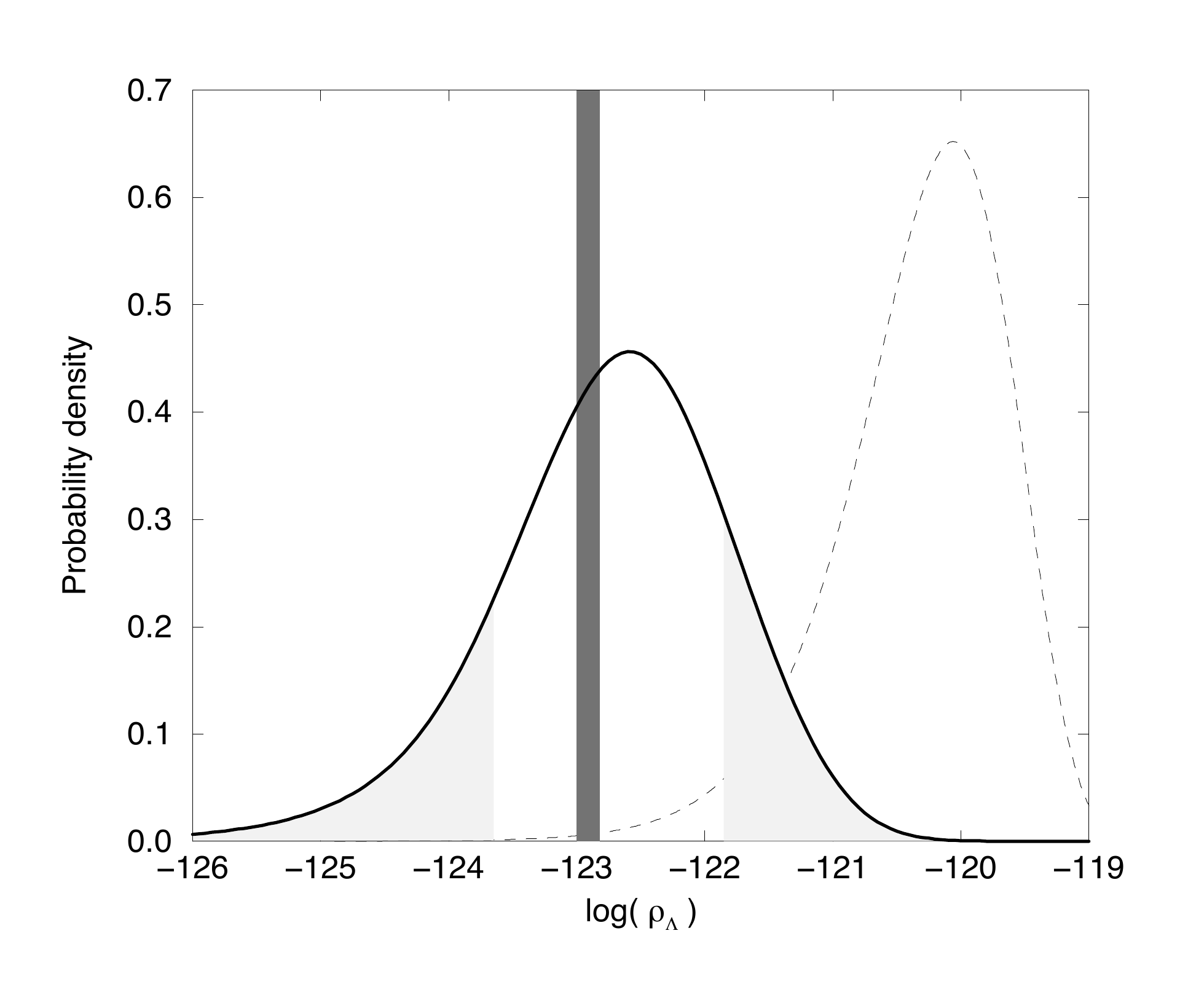}
 \caption{The vertical bar indicates the observed amount of vacuum energy (``dark energy'').  The solid line shows the prediction from the causal patch measure applied to the landscape of string theory, with the central 1$\sigma$ region indicated.  This plot is from Ref.~\cite{BouHar07}.  The agreement remains good independently of any assumptions about the nature of the observers.  The only relevant input parameter is the time when the observers emerge, $t_{\rm obs}\approx 13.7$ Gyr.}
\label{fig-stars_PS_hopkins} 
\end{figure}
Consider a class of observers that live at the time $t_{\rm obs}$ after the nucleation of their bubble universe.  Restricting attention to positive values of $\Lambda$, the causal patch measure~\cite{Bou06} predicts that such observers will find a cosmological constant
\begin{equation}
\Lambda\sim t_{\rm obs}^{-2}~.
\end{equation}
Using the observed value for the age of the universe, $t_{\rm obs}\approx 13.7$ Gyr, this result is in excellent agreement with observed value for the cosmological constant (see Fig.~\ref{fig-stars_PS_hopkins}).  

The successful prediction (or postdiction, in this case) of $\Lambda$ obtains independently of the nature of the observers.  For example, it applies to observers that do not require galaxies and even in vacua with very different low-energy physics.  In addition to the cosmological constant problem, it also addresses the coincidence problem discussed in Sec.~\ref{sec-coincidence}, since it predicts that observers should find themselves at the onset of vacuum domination, $t_\Lambda \sim t_{\rm obs}$.  Thus the prediction is more robust, and quantitatively more successful, than the seminal arguments of Weinberg~\cite{Wei87} and other early arguments requiring specific assumptions about observers~\cite{DavUnw82,Ban85,BarTip,Efs95}.  (The dashed line in Fig.~\ref{fig-stars_PS_hopkins} shows the prediction from the assumption that observers require galaxies, with an earlier measure developed in Ref.~\cite{MarSha97}.

There are currently no fully satisfactory measures for regions with nonpositive cosmological constant~\cite{BouFre10c}.  This remains a major outstanding challenge.  More broadly, it will be important to establish a solid theoretical basis for understanding both the landscape of string theory and the measure problem of eternal inflation.

\subsection{Connecting with Standard Cosmology}

How is the picture of a multiverse compatible with the one universe we see?  The multiverse is quite irregular, with different vacua in different places.  This appears to conflict with the observed homogeneity and isotropy of the visible universe.  We have not detected any other pocket universes.  As far as we can see, the vacuum seems to be the same, with the same particles, forces, and coupling constants.   Another concern is the claimed metastability of vacua.  If vacua can decay, how come our own vacuum is still around after billions of years?

In fact, all of these observations are generic predictions of the model, and all arise from the fact that vacuum decay is an exponentially suppressed tunneling effect.  This has three important consequences:
\begin{itemize} 
\item Individual pocket universes, including ours, can have very long lifetimes easily exceeding 10 Gyr~\cite{BP}.  
\item When a bubble of new vacuum does form, it will be highly symmetric~\cite{Col77}.  The symmetry of the decay process translates into the prediction that each pocket universe is a negatively curved, spatially homogeneous and isotropic universe~\cite{CDL}.
(The spatial curvature radius can be made unobservably large, as usual, by a period of slow-roll inflation at early times in our own pocket universe.)
\item Our parent vacuum need not produce many bubbles that collide with ours.  For such collisions to be visible, they would have to occur in our past light-cone, and the expected number of collisions can be $\ll 1$ for natural parameters.  
\end{itemize} 
Thus, the fact that we observe only one vacuum is not in contradiction with the string landscape.

However, this does not mean that other vacua will never be observed.  We would have to be somewhat lucky to observe a smoking gun signal of bubble collisions in the sky~\cite{FreKle09,CzeKle10,KleLev11,GobKle12}; for a review, see Ref.~\cite{Kle11}.  But it is a possibility, so the computation of its signature in the CMB for future searches such as PLANCK is of great interest~\cite{FeeJoh10a,FeeJoh10b,JohPei11,McEFee12}.  

Slow-roll inflation tends to wipe out signals from any era preceding it by stretching them to superhorizon scales.  If slow-roll inflation occurred after the formation of our bubble (as seems plausible), and if it lasted significantly longer than the 60 e-foldings necessary for explaining the observed flatness, then any imprints of bubble collisions or of  our parent vacuum will have been stretched to superhorizon scales.  

The decay of our own parent vacuum plays the role of what we used to call the big bang.  The vacuum energy of the parent vacuum is converted in part to the energy of the expanding domain wall bubble that separates our pocket universe from the parent vacuum.  But some of this energy can be dissipated later, inside our pocket universe.  It can drive a period of slow-roll inflation followed by the production of radiation and matter.

The decay of our parent vacuum will have taken place in an empty de~Sitter environment, so all matter and radiation in our vacuum must come from the vacuum energy released in the decay. In order to connect with standard cosmology, the energy density of radiation produced must be at least sufficient for nucleosynthesis. This constrains the vacuum energy of our parent vacuum: 
\begin{equation}
\Lambda_{\rm parent}\gg 10^{-88}~.
\label{eq-parent}
\end{equation} 

This constraint is very powerful.  Historically, it has ruled out one-dimensional potential landscapes such as the Abbott~\cite{Abb85} or Brown-Teitelboim~\cite{BT1,BT2} models, which were explicitly invented for the purpose of solving the cosmological constant problem.  In such models, neighboring vacua have nearly identical vacuum energy, $\Delta\Lambda<10^{-123}$.  Each decay lowers $\Lambda$ by an amount less than the observed value, so a very dense spectrum of $\Lambda$ is scanned over time.  This eventually produces a universe with $\Lambda$ as small as the observed value.  But because Eq.~(\ref{eq-parent}) is not satisfied, the universe is predicted to be empty, in conflict with observation.  One could invent one-dimensional landscapes in which the vacuum energy is random, but in natural models decay paths would end in terminal vacua with $\Lambda<0$ before reaching one of the rare vacua with $\Lambda\ll 1$.

In the string landscape, neighboring vacua typically have vastly different vacuum energy, with $\Lambda$ differing by as much as O(1) in Planck units (Sec.~\ref{sec-spectrum}).  Thus, matter and radiation can be produced in the decay of our parent vacuum.  Because the landscape is high-dimensional, there are many decay paths around terminal vacua.  Thus, all de~Sitter vacua in the landscape can be cosmologically produced by eternal inflation from generic initial conditions.   

It is interesting that string theory, which was not invented for the purpose of solving the cosmological constant problem, thus evades a longstanding obstruction.

\bibliographystyle{utcaps}
\bibliography{all}

\providecommand{\href}[2]{#2}\begingroup\raggedright\begin{thebibliography}{10}

\bibitem{Bou06b}
R.~Bousso, ``Precision cosmology and the landscape,''
\href{http://arxiv.org/abs/hep-th/0610211}{{\tt hep-th/0610211}}.
%%CITATION = HEP-TH 0610211;%%.

\bibitem{TASI07}
R.~Bousso, ``{TASI} Lectures on the Cosmological Constant,''
\href{http://arxiv.org/abs/arXiv:0708.4231 [hep-th]}{{\tt arXiv:0708.4231
  [hep-th]}}.
%%CITATION = ARXIV:0708.4231;%%.

\bibitem{HawEll}
S.~W. Hawking and G.~F.~R. Ellis, {\em The large scale stucture of space-time}.
\newblock Cambridge University Press, Cambridge, England, 1973.

\bibitem{GibHaw77a}
G.~W. Gibbons and S.~W. Hawking, ``Cosmological Event Horizons, Thermodynamics,
  and Particle Creation,'' {\em Phys. Rev. D} {\bf 15} (1977)  2738--2751.

\bibitem{Bou00a}
R.~Bousso, ``Positive vacuum energy and the {N}-bound,'' {\em JHEP} {\bf 11}
  (2000)  038,
\href{http://arxiv.org/abs/hep-th/0010252}{{\tt hep-th/0010252}}.
%%CITATION = HEP-TH 0010252;%%.

\bibitem{Edw72}
D.~Edwards, ``Exact expressions for the properties of the zero-pressure
  {F}riedmann models,'' {\em M.N.R.A.S.} {\bf 159} (1972)  51--66.

\bibitem{Wei89}
S.~Weinberg, ``The Cosmological Constant Problem,''
{\em Rev. Mod. Phys.} {\bf 61} (1989)  1--23.
%%CITATION = RMPHA,61,1;%%.

\bibitem{Pol06}
J.~Polchinski, ``The cosmological constant and the string landscape,''
\href{http://arxiv.org/abs/hep-th/0603249}{{\tt hep-th/0603249}}.
%%CITATION = HEP-TH 0603249;%%.

\bibitem{Zee}
A.~Zee, ``{Quantum field theory in a nutshell},''. ISBN-9780691140346.

\bibitem{Rie98}
{\bf Supernova Search Team} Collaboration, A.~G. Riess {\em et al.},
  ``Observational Evidence from Supernovae for an Accelerating Universe and a
  Cosmological Constant,'' {\em Astron. J.} {\bf 116} (1998)  1009--1038,
\href{http://arXiv.org/abs/astro-ph/9805201}{{\tt astro-ph/9805201}}.
%%CITATION = ASTRO-PH 9805201;%%.

\bibitem{Per98}
{\bf Supernova Cosmology Project} Collaboration, S.~Perlmutter {\em et al.},
  ``Measurements of {O}mega and {L}ambda from 42 High-Redshift Supernovae,''
  {\em Astrophys. J.} {\bf 517} (1999)  565--586,
\href{http://arXiv.org/abs/astro-ph/9812133}{{\tt astro-ph/9812133}}.
%%CITATION = ASTRO-PH 9812133;%%.

\bibitem{WMAP7}
{\bf WMAP} Collaboration, E.~Komatsu {\em et al.}, ``{Seven-Year Wilkinson
  Microwave Anisotropy Probe (WMAP) Observations: Cosmological
  Interpretation},'' \href{http://dx.doi.org/10.1088/0067-0049/192/2/18}{{\em
  Astrophys. J. Suppl.} {\bf 192} (2011)  18},
\href{http://arxiv.org/abs/1001.4538}{{\tt arXiv:1001.4538 [astro-ph.CO]}}.
%%CITATION = 1001.4538;%%.

\bibitem{Rei09}
B.~A. Reid {\em et al.}, ``{Cosmological Constraints from the Clustering of the
  Sloan Digital Sky Survey DR7 Luminous Red Galaxies},''
  \href{http://dx.doi.org/10.1111/j.1365-2966.2010.16276.x}{{\em Mon. Not. Roy.
  Astron. Soc.} {\bf 404} (2010)  60--85},
\href{http://arxiv.org/abs/0907.1659}{{\tt arXiv:0907.1659 [astro-ph.CO]}}.
%%CITATION = 0907.1659;%%.

\bibitem{Car00}
S.~M. Carroll, ``The cosmological constant,''
\href{http://arxiv.org/abs/astro-ph/0004075}{{\tt astro-ph/0004075}}.
%%CITATION = ASTRO-PH 0004075;%%.

\bibitem{Car98}
S.~M. Carroll, ``{Quintessence and the rest of the world},''
  \href{http://dx.doi.org/10.1103/PhysRevLett.81.3067}{{\em Phys. Rev. Lett.}
  {\bf 81} (1998)  3067--3070},
\href{http://arxiv.org/abs/astro-ph/9806099}{{\tt arXiv:astro-ph/9806099}}.
%%CITATION = ASTRO-PH/9806099;%%.

\bibitem{HalNom05}
L.~J. Hall, Y.~Nomura, and S.~J. Oliver, ``{Evolving dark energy with w not
  equal -1},'' \href{http://dx.doi.org/10.1103/PhysRevLett.95.141302}{{\em
  Phys. Rev. Lett.} {\bf 95} (2005)  141302},
\href{http://arxiv.org/abs/astro-ph/0503706}{{\tt arXiv:astro-ph/0503706}}.
%%CITATION = ASTRO-PH/0503706;%%.

\bibitem{DETF}
A.~Albrecht {\em et al.}, ``Report of the {D}ark {E}nergy {T}ask {F}orce,''
\href{http://arxiv.org/abs/astro-ph/0609591}{{\tt astro-ph/0609591}}.
%%CITATION = ASTRO-PH/0609591;%%.

\bibitem{Sak84}
A.~D. Sakharov, ``Cosmological Transitions with a Change in Metric Signature,''
{\em Sov. Phys. JETP} {\bf 60} (1984)  214--218.
%%CITATION = SPHJA,60,214;%%.

\bibitem{GSW}
M.~B. Green, J.~H. Schwarz, and E.~Witten, {\em Superstring {T}heory}.
\newblock Cambridge Univ. Pr., Cambridge, UK, 1987.

\bibitem{Polchinski}
J.~Polchinski, {\em String {T}heory}.
\newblock Cambridge Univ. Pr., Cambridge, UK, 1998.

\bibitem{BouPol04}
R.~Bousso and J.~Polchinski, ``{The string theory landscape},''
{\em Sci. Am.} {\bf 291} (2004)  60--69.
%%CITATION = SCAMA,291,60;%%.

\bibitem{LerLus87}
W.~Lerche, D.~Lust, and A.~N. Schellekens, ``{Chiral Four-Dimensional Heterotic
  Strings from Selfdual Lattices},''
\href{http://dx.doi.org/10.1016/0550-3213(87)90115-5}{{\em Nucl. Phys.} {\bf
  B287} (1987)  477}.
%%CITATION = NUPHA,B287,477;%%.

\bibitem{DouKac06}
M.~R. Douglas and S.~Kachru, ``Flux compactification,''
\href{http://arxiv.org/abs/hep-th/0610102}{{\tt hep-th/0610102}}.
%%CITATION = HEP-TH 0610102;%%.

\bibitem{Sch98}
A.~N. Schellekens, ``{The landscape 'avant la lettre'},''
\href{http://arxiv.org/abs/physics/0604134}{{\tt arXiv:physics/0604134}}.
%%CITATION = PHYSICS/0604134;%%.

\bibitem{BP}
R.~Bousso and J.~Polchinski, ``Quantization of four-form fluxes and dynamical
  neutralization of the cosmological constant,'' {\em JHEP} {\bf 06} (2000)
  006,
\href{http://arxiv.org/abs/hep-th/0004134}{{\tt hep-th/0004134}}.
%%CITATION = JHEPA,0006,006;%%.

\bibitem{KKLT}
S.~Kachru, R.~Kallosh, A.~Linde, and S.~P. Trivedi, ``De {S}itter vacua in
  string theory,'' {\em Phys. Rev. D} {\bf 68} (2003)  046005,
\href{http://arxiv.org/abs/hep-th/0301240}{{\tt hep-th/0301240}}.
%%CITATION = HEP-TH 0301240;%%.

\bibitem{Sil01}
E.~Silverstein, ``{(A)dS} backgrounds from asymmetric orientifolds,''
\href{http://arXiv.org/abs/hep-th/0106209}{{\tt hep-th/0106209}}.
%%CITATION = HEP-TH 0106209;%%.

\bibitem{MalSil02}
A.~Maloney, E.~Silverstein, and A.~Strominger, ``De {S}itter space in
  noncritical string theory,''
\href{http://arxiv.org/abs/hep-th/0205316}{{\tt hep-th/0205316}}.
%%CITATION = HEP-TH 0205316;%%.

\bibitem{DenDou04b}
F.~Denef and M.~R. Douglas, ``Distributions of flux vacua,'' {\em JHEP} {\bf
  05} (2004)  072,
\href{http://arxiv.org/abs/hep-th/0404116}{{\tt hep-th/0404116}}.
%%CITATION = HEP-TH 0404116;%%.

\bibitem{Col77}
S.~Coleman, ``The Fate of the False Vacuum. 1. {S}emiclassical Theory,'' {\em
  Phys. Rev. D} {\bf 15} (1977)  2929--2936.

\bibitem{CDL}
S.~Coleman and F.~D. Luccia, ``Gravitational effects on and of vacuum decay,''
  {\em Phys. Rev. D} {\bf 21} (1980)  3305--3315.

\bibitem{BroDah10}
A.~R. Brown and A.~Dahlen, ``{Giant Leaps and Minimal Branes in
  Multi-Dimensional Flux Landscapes},''
  \href{http://dx.doi.org/10.1103/PhysRevD.84.023513}{{\em Phys. Rev.} {\bf
  D84} (2011)  023513},
\href{http://arxiv.org/abs/1010.5241}{{\tt arXiv:1010.5241 [hep-th]}}.
%%CITATION = 1010.5241;%%.

\bibitem{GutWei83}
A.~H. Guth and E.~J. Weinberg, ``Could the universe have recovered from a slow
  first-order phase transition?,'' {\em Nucl. Phys.} {\bf B212} (1983)
  321--364.

\bibitem{Lin86a}
A.~Linde, ``Eternally Existing Selfreproducing Chaotic Inflationary Universe,''
  {\em Phys. Lett.} {\bf B175} (1986)  395--400.

\bibitem{BouFre10a}
R.~Bousso, B.~Freivogel, and S.~Leichenauer, ``{Saturating the holographic
  entropy bound},'' \href{http://dx.doi.org/10.1103/PhysRevD.82.084024}{{\em
  Phys. Rev.} {\bf D82} (2010)  084024},
\href{http://arxiv.org/abs/1003.3012}{{\tt arXiv:1003.3012 [hep-th]}}.
%%CITATION = 1003.3012;%%.

\bibitem{CEB1}
R.~Bousso, ``A covariant entropy conjecture,'' {\em JHEP} {\bf 07} (1999)  004,
\href{http://arxiv.org/abs/hep-th/9905177}{{\tt hep-th/9905177}}.
%%CITATION = JHEPA,9907,004;%%.

\bibitem{RMP}
R.~Bousso, ``The holographic principle,'' {\em Rev. Mod. Phys.} {\bf 74} (2002)
   825,
\href{http://arXiv.org/abs/hep-th/0203101}{{\tt hep-th/0203101}}.
%%CITATION = HEP-TH 0203101;%%.

\bibitem{BouFre10d}
R.~Bousso, B.~Freivogel, S.~Leichenauer, and V.~Rosenhaus, ``{A geometric
  solution to the coincidence problem, and the size of the landscape as the
  origin of hierarchy},''
  \href{http://dx.doi.org/10.1103/PhysRevLett.106.101301}{{\em Phys. Rev.
  Lett.} {\bf 106} (2011)  101301},
\href{http://arxiv.org/abs/1011.0714}{{\tt arXiv:1011.0714 [hep-th]}}.
%%CITATION = 1011.0714;%%.

\bibitem{BouHar07}
R.~Bousso, R.~Harnik, G.~D. Kribs, and G.~Perez, ``Predicting the cosmological
  constant from the causal entropic principle,'' {\em Phys. Rev. D} {\bf 76}
  (2007)  043513,
\href{http://arxiv.org/abs/hep-th/0702115}{{\tt hep-th/0702115}}.
%%CITATION = HEP-TH 0702115;%%.

\bibitem{BouHar10}
R.~Bousso and R.~Harnik, ``{The Entropic Landscape},''
  \href{http://dx.doi.org/10.1103/PhysRevD.82.123523}{{\em Phys. Rev.} {\bf
  D82} (2010)  123523},
\href{http://arxiv.org/abs/1001.1155}{{\tt arXiv:1001.1155 [hep-th]}}.
%%CITATION = 1001.1155;%%.

\bibitem{Bou06}
R.~Bousso, ``Holographic probabilities in eternal inflation,'' {\em Phys. Rev.
  Lett.} {\bf 97} (2006)  191302,
\href{http://arxiv.org/abs/hep-th/0605263}{{\tt hep-th/0605263}}.
%%CITATION = HEP-TH/0605263;%%.

\bibitem{Wei87}
S.~Weinberg, ``Anthropic Bound on the Cosmological Constant,''
{\em Phys. Rev. Lett.} {\bf 59} (1987)  2607.
%%CITATION = PRLTA,59,2607;%%.

\bibitem{DavUnw82}
P.~C.~W. Davies and S.~D. Unwin, ``Why is the cosmological constant so small,''
{\em Proc. R. Soc. Lond., A} {\bf 377} (1982)  147--149.
%%CITATION = PHRVA,D25,942;%%.

\bibitem{Ban85}
T.~Banks, ``{T C P}, Quantum Gravity, the Cosmological Constant and All That,''
{\em Nucl. Phys.} {\bf B249} (1985)  332.
%%CITATION = NUPHA,B249,332;%%.

\bibitem{BarTip}
J.~D. Barrow and F.~J. Tipler, {\em The Anthropic Cosmological Principle}.
\newblock Clarendon Press, Oxford, 1986.

\bibitem{Efs95}
G.~P. Efstathiou, ``An anthropic argument for a cosmological constant,'' {\em
  M.N.R.A.S.} {\bf 274} (1995)  L73.

\bibitem{MarSha97}
H.~Martel, P.~R. Shapiro, and S.~Weinberg, ``Likely Values of the Cosmological
  Constant,''
\href{http://arxiv.org/abs/astro-ph/9701099}{{\tt astro-ph/9701099}}.
%%CITATION = ASTRO-PH 9701099;%%.

\bibitem{BouFre10c}
R.~Bousso, B.~Freivogel, S.~Leichenauer, and V.~Rosenhaus, ``{Eternal inflation
  predicts that time will end},''
  \href{http://dx.doi.org/10.1103/PhysRevD.83.023525}{{\em Phys. Rev.} {\bf
  D83} (2011)  023525},
\href{http://arxiv.org/abs/1009.4698}{{\tt arXiv:1009.4698 [hep-th]}}.
%%CITATION = 1009.4698;%%.

\bibitem{FreKle09}
B.~Freivogel, M.~Kleban, A.~Nicolis, and K.~Sigurdson, ``{Eternal Inflation,
  Bubble Collisions, and the Disintegration of the Persistence of Memory},''
\href{http://arxiv.org/abs/0901.0007}{{\tt arXiv:0901.0007 [hep-th]}}.
%%CITATION = 0901.0007;%%.

\bibitem{CzeKle10}
B.~Czech, M.~Kleban, K.~Larjo, T.~S. Levi, and K.~Sigurdson, ``{Polarizing
  Bubble Collisions},''
  \href{http://dx.doi.org/10.1088/1475-7516/2010/12/023}{{\em JCAP} {\bf 1012}
  (2010)  023},
\href{http://arxiv.org/abs/1006.0832}{{\tt arXiv:1006.0832 [astro-ph.CO]}}.
%%CITATION = 1006.0832;%%.

\bibitem{KleLev11}
M.~Kleban, T.~S. Levi, and K.~Sigurdson, ``{Observing the Multiverse with
  Cosmic Wakes},''
\href{http://arxiv.org/abs/1109.3473}{{\tt arXiv:1109.3473 [astro-ph.CO]}}.
%%CITATION = 1109.3473;%%.

\bibitem{GobKle12}
R.~Gobbetti and M.~Kleban, ``{Analyzing Cosmic Bubble Collisions},''
\href{http://arxiv.org/abs/1201.6380}{{\tt arXiv:1201.6380 [hep-th]}}.
%%CITATION = 1201.6380;%%.

\bibitem{Kle11}
M.~Kleban, ``{Cosmic Bubble Collisions},''
  \href{http://dx.doi.org/10.1088/0264-9381/28/20/204008}{{\em Class. Quant.
  Grav.} {\bf 28} (2011)  204008},
\href{http://arxiv.org/abs/1107.2593}{{\tt arXiv:1107.2593 [astro-ph.CO]}}.
%%CITATION = 1107.2593;%%.

\bibitem{FeeJoh10a}
S.~M. Feeney, M.~C. Johnson, D.~J. Mortlock, and H.~V. Peiris, ``{First
  Observational Tests of Eternal Inflation},''
  \href{http://dx.doi.org/10.1103/PhysRevLett.107.071301}{{\em Phys. Rev.
  Lett.} {\bf 107} (2011)  071301},
\href{http://arxiv.org/abs/1012.1995}{{\tt arXiv:1012.1995 [astro-ph.CO]}}.
%%CITATION = 1012.1995;%%.

\bibitem{FeeJoh10b}
S.~M. Feeney, M.~C. Johnson, D.~J. Mortlock, and H.~V. Peiris, ``{First
  Observational Tests of Eternal Inflation: Analysis Methods and WMAP 7-Year
  Results},'' \href{http://dx.doi.org/10.1103/PhysRevD.84.043507}{{\em Phys.
  Rev.} {\bf D84} (2011)  043507},
\href{http://arxiv.org/abs/1012.3667}{{\tt arXiv:1012.3667 [astro-ph.CO]}}.
%%CITATION = 1012.3667;%%.

\bibitem{JohPei11}
M.~C. Johnson, H.~V. Peiris, and L.~Lehner, ``{Determining the outcome of
  cosmic bubble collisions in full General Relativity},''
\href{http://arxiv.org/abs/1112.4487}{{\tt arXiv:1112.4487 [hep-th]}}.
%%CITATION = 1112.4487;%%.

\bibitem{McEFee12}
J.~D. McEwen, S.~M. Feeney, M.~C. Johnson, and H.~V. Peiris, ``{Optimal filters
  for detecting cosmic bubble collisions},''
\href{http://arxiv.org/abs/1202.2861}{{\tt arXiv:1202.2861 [astro-ph.CO]}}.
%%CITATION = 1202.2861;%%.

\bibitem{Abb85}
L.~F. Abbott, ``A Mechanism for Reducing the Value of the Cosmological
  Constant,''
{\em Phys. Lett.} {\bf B150} (1985)  427.
%%CITATION = PHLTA,B150,427;%%.

\bibitem{BT1}
J.~D. Brown and C.~Teitelboim, ``Dynamical Neutralization of the Cosmological
  Constant,'' {\em Phys. Lett.} {\bf B195} (1987)  177.

\bibitem{BT2}
J.~D. Brown and C.~Teitelboim, ``Neutralization of the Cosmological Constant by
  Membrane Creation,'' {\em Nucl. Phys.} {\bf B297} (1988)  787.

\end{thebibliography}\endgroup
\end{document}